\documentstyle[preprint,pre,aps,eqsecnum]{revtex}
\tightenlines

\begin{document}

\draft
\title{SIDEBRANCHING INDUCED BY EXTERNAL NOISE IN SOLUTAL DENDRITIC
GROWTH}

\author{R. Gonz\'alez-Cinca$^{1,*}$, L. Ram\'{\i}rez-Piscina$^1$,
J.Casademunt$^2$, A. Hern\'{a}ndez-Machado$^2$}

\address{
$^1$ Departament de F\'{\i}sica Aplicada,
Universitat Polit\`ecnica de Catalunya,
Campus Nord UPC, M\`odul B5,
E-08034 Barcelona, SPAIN \\
$^2$ Departament E.C.M., Facultat de F\'{\i}sica,
Universitat de Barcelona,
Diagonal 647, E-08028 Barcelona, SPAIN\\
$*$ Corresponding author. Tel.: (+34) 934017983, fax: (+34)
934016090,  e-mail: ricard@fa.upc.es
}

\date{\today}

\maketitle

\begin{abstract}

We have studied  sidebranching induced by fluctuations in dendritic growth.
The amplitude of sidebranching induced by internal
(equilibrium) concentration fluctuations in the case of
solidification with solutal diffusion is computed. This amplitude turns out  to be
significantly smaller than values reported in previous
experiments.The effects of other possible sources of fluctuations
(of an external origin) are examined by introducing  non-conserved noise in a
phase-field model. This reproduces the characteristics of
sidebranching found in experiments. Results also show that
sidebranching induced by external noise is qualitatively
similar to that of internal noise, and it is only distinguished
by its amplitude.

\end{abstract}

\pacs{PACS numbers: 68.70.+w, 81.10.Aj, 64.70.Dv, 05.40.-a, 81.30.Fb, 05.70.Ln}


\section{Introduction}

Dendritic growth in nonequilibrium systems has been extensively
studied during  the last few years
\cite{langer,pelce,benjacob,brener,godreche,hurle,santafe}.
A feature  which remains a main point of interest is the study of
sidebranching, which is the secondary branches that appear at both
sides of the main dendrite. The question of how its frequency and
amplitude are determined has not yet been fully solved.
Two scenarios have been proposed to explain the origin of
sidebranching. One of them states that periodic deterministic
oscillations at the tip \cite{langer2,martin,lacombe} generate correlated branches on both flanks
of the dendrite \cite{langer2,martin}.
A possible source of these tip oscillations was suggested
in Ref. \cite{vansaarloos}, where it was argued that the spreading
rate of the wavepacket that characterizes sidebranching might
become large enough so that the tip could undergo oscillations or 
other instabilities. This is predicted to occur in the limit of
small surface tension anisotropy. The
other scenario proposes that sidebranching is due to selective
amplification of fluctuations
near the tip of the dendrite
\cite{pieters,kessler,barber,pieters2,langer3,brener2,dougherty,qian,bouissou,bisang1,bisang2,hurlimann,williams,pavlik,karma2,pavlik2}.
In this case, branches appear to be mostly uncorrelated. In this paper we
will study this second scenario by means of a phase-field model
\cite{pavlik,karma2,pavlik2,kobayashi,wheeler,mcfadden,wang,braun,murray,wang2,physd,pre,karma,jcg,tamas,tamas2},
and, in particular, we will focus on the issue of
an external vs. internal origin of
the noise.

In a frame of reference moving with the tip of the dendrite,
sidebranching can be seen as a wave that propagates along the
dendrite away from the tip at the same velocity as the tip. An appropriate
characterization is provided by the  amplitude and
wavelength.
Barber et al. \cite{barber} studied the
evolution of time-dependent  deformations of the needle crystal
(Ivantsov) solution of the two-dimensional symmetric model of
solidification in the limit of small P\`eclet number within a WKB approximation. The amplitude
of a localized wave packet grows exponentially as $z^{1/4}$, where
$z$ is measured from the tip along the symmetry axis of the
dendrite, as the packet moves down, provided the initial packet
contains modes of arbitrarily small frequencies. Moreover, the
wavelength of the packet increases as $z^{1/4}$.
Pieters \cite{pieters2} obtained the same 
amplitude and wavelength  dependence on $z$ as in Ref. \cite{barber} both
analytically and by numerical integration of the boundary-layer
model. Langer \cite{langer3} concluded, from a similar analysis to
that of Ref. \cite{barber} but performed in three dimensions, that
noise of some kind can be the origin of sidebranching, but that
thermal fluctuations are not strong enough to entirely explain the
phenomena. However, more recently, Brener and Temkin \cite{brener2}
considered anisotropic needle crystals in three dimensions and
concluded that experimentally observed sidebranching could be
explained by considering noise of a thermal origin. The growth of the
sidebranching amplitude was found to behave exponentially as a
function of $z^{2/5}$, which is faster than the $z^{1/4}$
dependence obtained in the axisymmetric case \cite{langer3}. The
sidebranching wavelength was found to be a function  of $z^{1/5}$,
very similar to that obtained in the axisymmetric case.
Dougherty et al. \cite{dougherty} studied sidebranching in
$NH_{4}Br$ dendrites, where rather uncorrelated variations in phase
and amplitude of the branches were observed. They determined the
amplitude of the sidebranching and its mean wavelength
by looking at the power spectrum of the data obtained by measuring
the half-width of the dendrite at a fixed distance $z$ from the tip
at different times. The behaviour of the amplitude was qualitatively
like that predicted in Ref. \cite{barber} up to a certain value of
$z$, after which the linear theory is presumably no longer valid. An
equivalent exponential growth of the amplitude with $s^{1/4}$,
where $s$ is the arclength, was also found in Ref. \cite{qian}.
However, in Ref. \cite{dougherty} no variation of the mean
frequency in the spectral peak was obtained for different $z$.
Finally, Dougherty et al. also observed that side branches
separated by more than about six times the mean wavelength were
uncorrelated. Weak correlation between opposite sides of the
dendrite when no external forcing was applied to it was also found
in Ref. \cite{bouissou}. The common feature of all these
experiments is that sidebranching appears to be due to the selective
amplification of natural noise and not to the existence of some
intrinsic oscillation or limit cycle.

Bisang and Bilgram \cite{bisang1,bisang2} found  quantitative
agreement between the predictions for the linear regime in
\cite{brener2} and their results in experiments on xenon
dendrites in three dimensions. They concluded that Brener and
Temkin's theory correctly describes the sidebranching behaviour of
dendrites for any pure substance  with cubic symmetry and thus
thermal noise was concluded to be
the origin of the sidebranching observed in their
experiments.

In the last decade there has been an increasing use of phase-field
models to deal with dendritic growth problems.
They are a very useful and practical
tool to simulate such kind of processes and a good alternative to
the integrodifferential equation which can be derived from the
classical sharp-interface model. In the phase-field models an
order parameter or phase field $\phi$ is defined, which avoids
tracking the interface and naturally includes the physical
boundary conditions at the interface.

Up to now, few studies of sidebranching phenomena with the
phase-field model have been carried out. It has been shown that
the inclusion of a noise source in the phase-field model
equations enhances the emergence of uncorrelated sidebranching
\cite{kobayashi} without
affecting the velocity and radius of the tip \cite{wheeler}.
Moreover, when the dendrite tip is periodically forced, the
sidebranching appears to be correlated at both sides of the dendrite
\cite{murray}  as has been observed in some experiments
\cite{bouissou,williams}. In particular, sidebranching can be
regulated by
spatially homogeneous time-periodic variation of the melting point
induced by oscillations in the external pressure or by periodic
heating generated by a dissipative electric current \cite{tamas2}.

The deepest insight into the study of sidebranching with a
phase-field model have been carried out recently in Ref.
\cite{karma2}. They included thermal noise in a
two-dimensional phase-field model of solidification controlled by
heat-diffusion in a way which was consistent with both bulk and interfacial
equilibrium fluctuations, as has been done previously with the
sharp-interface model equations \cite{karma3,karma4}. Karma and
Rappel \cite{karma2} obtained good quantitative agreement between
the computed sidebranching amplitude as a function
of distance to the tip and the prediction of the linear $WKB$
theory for anisotropic crystals in two dimensions. 
Sidebranching wavelength very close to the tip was found to increase with
$z$ faster than predicted by the $WKB$ theory, but this could be explained
after considering that perturbations generally get stretched as
they travel along the sides of curved fronts. Further from the tip, the value
of the sidebranching
wavelength saturates.

Although there is general agreement in that thermal (internal)
fluctuations are enough to explain the amplitude of the dendritic
sidebranching, one should be aware that evidence along these lines has been achieved in experiments of heat-controlled
solidification of pure substances. As up to now there is a lack of
predictions of sidebranching amplitudes for solutal dendrites,
experiments of these dendrites can only show qualitative
agreement with theoretical results.
In this paper we address the question of sidebranching
characteristics in the presence of external vs. internal
fluctuations. First of all, we obtain a prediction of the effects
of internal noise on sidebranching amplitudes for solutal
dendrites. A comparison of the theory with available quantitative
experimental results \cite{dougherty} shows that there are serious indications that
in some experiments internal thermodynamical fluctuations could not
account for observed sidebranching activity. In this case some
other source of fluctuations, of an external origin, should be
called on. Some of the consequences derived from adding
a non-conserved noise source in a  two-dimensional phase-field
model are examined. This noise is of a different nature to what one should
employ to provide an account of internal fluctuations
\cite{pavlik,karma2,pavlik2,karma3,karma4}. However, our simulations qualitatively
reproduce the properties of the noise-induced sidebranching derived
analytically \cite{brener2,karma2} and observed experimentally
\cite{dougherty,qian}.
We conclude that although thermal noise is not always the main
origin of dendritic sidebranching, its qualitative characteristics
are common for noise-induced sidebranching independently of its
origin. A detailed quantitative study of sidebranching activity
could therefore be useful to elucidate the origin of the noise
in specific experiments.

In Section II we predict the sidebranching amplitudes for solutal
dendrites with thermal fluctuations.
In Section III the model and the numerical method used to solve the
equations are described. Numerical results on the behaviour of the
sidebranching characteristics as well as comparison with
theoretical predictions and experimental results are described in
Section IV. Conclusions derived from these results are outlined in
Section V.

\section{Sidebranching amplitude in solutal solidification}

Available theoretical predictions on sidebranching amplitudes have
been formulated for dendrites grown from a pure substance and
controlled by heat diffusion.
Here we will consider dendrites appearing in isothermal growth of
mixtures controlled by diffusion of the solute. We start from the
Langevin formalism for solidification due to Karma
\cite{karma3,karma4}. In this formalism the usual sharp-interface
model for solidification is completed with noise terms constructed
with the requirement that they give the correct bulk and interfacial
equilibrium fluctuations.
The resulting diffusion equation for a mixture in isothermal conditions is:
\begin{eqnarray}
\label{diffusion}
\frac{\partial C_\nu }{\partial t}=D_\nu \Delta C_\nu -\nabla \cdot
{\bf q}^\nu ({\bf r},t),
\end{eqnarray}
with the following boundary conditions at the interface:
\begin{equation}
\label{masscons}
(C_L-C_S)v_n=
{\bf \hat n}\cdot \left[ D_S\nabla C_S-D_L\nabla C_L\right] +{\bf
\hat n}
\cdot \left[ {\bf q}^L-{\bf q}^S\right]
\end{equation}
\begin{equation}
\label{gibbsthomson}
m_EC_L+\Gamma \kappa +\frac{v_n}
\mu =T_M-T+\eta ({\bf r},t),
\end{equation}
where $\nu =S,L$ denotes the phase, $C_\nu $ is the concentration,
$T_M-T$
is the undercooling, $D_\nu $ is the diffusion coefficient, $m_E$
is the absolute value of the (negative) $T(C_L)$ slope of the
coexistence curve, and $\kappa$, $\bf \hat n$ and $v_n$ are
the
curvature, the normal unitary vector and the normal velocity of the
interface, respectively. $\Gamma=\sigma T/L$, $\sigma$ is the surface
energy and $L$ the latent heat per unit volume. ${\bf q}$ and $\eta$
are random forces whose statistical properties are given by
\begin{eqnarray}
\label{intensq}
\left\langle q_i^\nu (
{\bf r},t)q_j^\nu ({\bf r}^{\prime },t^{\prime })\right\rangle
=2D_\nu
C_\nu ({\bf r},t)\delta ({\bf r,r}^{\prime })\delta (t-t^{\prime
})\delta _{ij}
\\ \label{intenseta}
\left\langle \eta _i^\nu ({\bf r}_{\perp },t)\eta _j^\nu (%
{\bf r}_{\perp }^{\prime },t^{\prime })\right\rangle
=2\frac{k_BT^2}{\mu L}%
\frac{\delta ({\bf r}_{\perp }{\bf ,r}_{\perp }^{\prime })\delta
(t-t^{\prime })}{\sqrt{1+\left| \nabla _{\perp }\xi ({\bf r}_{\perp
},t)\right| }},
\end{eqnarray}
where the interface is parametrized by the vector
${\bf r}_{\perp } + \xi {\bf \hat n}$.

These equations can be mapped under several approximations into the
corresponding Langevin model for free solidification of a pure
substance \cite{karma2,karma3,karma4}. First we assume a constant
concentration gap in the mass conservation Eq. \ref{masscons}, {\it
i.e.} $C_L-C_S \equiv \Delta C \approx \Delta C^{eq}$, the value
corresponding to equilibrium at temperature $T$.
This is in principle valid for small curvatures and velocities. A
similar approximation
is assumed in the intensity of the bulk noise $q_i^\nu ({\bf
r},t)$,
substituting $C_\nu ({\bf r},t)$ by the equilibrium value
$C_\nu^{eq}$ in
Eq. \ref{intensq}. Furthermore in the intensity of the
interfacial noise $\eta _i^\nu ({\bf r}_{\perp},t)$ we employ the
Clausius-Clapeyron equation for dilute alloys \cite{karma4} to
make the substitution $\frac{k_BT^2}{\mu L}\approx \frac{%
C_L^{eq}m_E}{\mu \Delta C^{eq}}$ in Eq. \ref{intenseta}.
Within these approximations the process of isothermal
solidification of an alloy is equivalent (including thermodynamical
fluctuations) to the (heat diffusion controlled) solidification of
a pure substance, whose specific heat, latent heat and melting
temperature are given by
\begin{eqnarray}
\label{map1}
c_p=k_B
\frac{\left( C_L^{eq}\right) ^3}{\left( \Delta C^{eq}\right) ^2} \\
\label{map2}
L=k_B
\frac{\left( C_L^{eq}\right) ^3}{\Delta C^{eq}}m_E \\
\label{map3}
{\overline T}_M=\frac{\left(
C_L^{eq}\right) ^2}{\Delta C^{eq}}m_E
\end{eqnarray}
where the diffusion field is now a temperature field given by
\begin{equation}
{\overline T}({\bf r},t)-{\overline T}_M=\left(
C_\nu-C_\nu^{eq}\right) m_E
\end{equation}
and the rest of parameters remain unchanged. This can be checked by
direct substitution in the Langevin equations. Therefore, the
sidebranching induced by thermodynamical fluctuations should be the
same in both situations.
Sidebranching amplitude is predicted to depend on the distance $z$
along the dendrite axis as \cite{brener2}:%
\begin{equation}
\label{sidebramp}
A(z)=\rho \overline{S}\exp \left( \frac 23\left[
\frac{x_0^{3}(z)}{3\sigma
^{*}z\rho ^2}\right] ^{1/2}\right),
\end{equation}
where $\rho $ is the tip radius, $x_0(z)$ is the shape of the
dendrite, and the operating mode of the dendrite is given by the
parameter $\sigma^{*}$ defined by
\begin{equation}
\sigma ^{*}=2D\ d_0/\rho ^2V,
\end{equation}
where $V$ is the selected velocity and $d_0$ is the capillary
length.
The dimensionless noise amplitude, for a $d$-dimensional thermal dendrite, is then
known to be
\cite{brener2,karma2}%
\begin{equation}
\overline{S}^2=\frac{2k_BT_M^2c_pD}{L^2\rho ^{1+d}V}.
\end{equation}
Applying the mapping above, the corresponding result for a $d$-dimensional solutal
dendrite is%
\begin{equation}
\label{sidebrnoise}
\overline{S}^2=\frac{2C_L^{eq}D}{\left( \Delta C^{eq}\right) ^2\rho
^{1+d}V}.
\end{equation}
This result will be used below to compare the prediction for the case
of internal noise with experiments on solutal dendrites, for which
there are not many quantitative experimental results available.
We focus on the experiments
performed in Ref. \cite{dougherty} with ammonium bromide dendrites
growing from supersaturated aqueous solution in isothermal
conditions. In this experiment the precipitate front
advances by incorporating solute particles instead of rejecting
them, which makes it differ by several details from standard
solutal solidification. Nevertheless, the above result (Eq. \ref{sidebrnoise}) can be
obtained by slightly adapting the performed mapping. In this case the
system is on the high concentration side of the phase diagram of the
mixture, for which $T_M$ is that of the solvent, no
longer close to the temperature of the experiment. In this case it
is convenient to write the Gibbs-Thomson equation as
\begin{equation}
\label{gibbsthomson2}
m_E({C_\infty} - {C_L})+\Gamma \kappa +\frac{v_n}
\mu =T_S-T+\eta ({\bf r},t),
\end{equation}
where $C_\infty$ is the concentration of the dilution, $T_S$ is the
saturation temperature for that concentration, and $m_E$ is now the
(positive) $T(C_L)$ slope. The same results of Eqs.
\ref{sidebramp} and \ref{sidebrnoise} are obtained by applying the
mapping of Eqs. \ref{map1}, \ref{map2} and \ref{map3} with a diffusion
field
\begin{equation}
{\overline T}({\bf r},t)-{\overline T}_M=\left(
C_\nu^{eq}-C_\nu\right) m_E,
\end{equation}
where we have used the relation
$T_S - T = m_E (C_\infty - C_L^{eq})$.

Now we compare the prediction of Eqs.
\ref{sidebramp} and \ref{sidebrnoise} with the experimental results on
supersaturated solutions of Ref. \cite{dougherty}. These
experiments were performed at a supersaturation $\Delta =
(C_{\infty}-C_{eq})/(C_{S}-C_{eq}) = 0.007$ and a saturation
temperature of $56^{o}C$. The characteristics of the selected
dendrite are $\rho=4.0 \mu m$, $V=1.44 \mu m/s$ and $\sigma^{*} =
0.081$. We have employed a value of $D=2.6 \times 10^{-5} cm^2 / s$
and the values corresponding to a temperature of $100^{o}C$ for the
equilibrium concentrations: $C_L^{eq} = 0.99 \times 10^4 N_A
molec/m^3$ and $C_S^{eq} = 2.48 \times 10^4 N_A molec/m^3$. Since
this temperature is much higher than that of the experiment, the
resulting value $\overline{S} = 5.96 \times 10^{-5}$ constitutes an
overestimation of the theoretical value.

In Fig. 1 we show the calculated amplitude $A(z)$ of the
sidebranching induced by internal fluctuations
for this value of $\overline S$. 
We consider the theoretically predicted shape
$x(z)=(\frac{5}{3} z)^{3/5}$, as was considered in Ref. \cite{brener2}.
The result corresponding to the actual
temperature of the experiment would be placed below the represented
curve.
In the same figure we plot
the experimental results of Ref.
\cite{dougherty}. 
We see that experimental amplitudes
are approximately one order of magnitude larger than the 
overestimated theoretical values.

Therefore 
the predicted amplitude of the
sidebranching when it is due to statistical noise  is,
for the experiments of Ref. \cite{dougherty},
at least one order of magnitude smaller than the amplitude
experimentally observed. Thus, we
are led to conclude that thermodynamical fluctuations are not
enough to explain the sidebranching amplitude in some experiments.
In this estimation we have assumed three-dimensional dendrites even
though the experiments of Ref. \cite{dougherty} are intended to be
quasi two-dimensional. The analogous calculations in two dimensions
show an even lower sidebranching amplitude, i.e. a larger discrepancy
with the experiments.

There is a shortcoming in the predictions above when
applied to supersaturated experiments. By their own nature,
supersaturated dilutions are not in the diluted limit, as assumed
in the theoretical analysis \cite{karma4}. In the experiments of Ref.
\cite{dougherty} the concentration is as high as $16\%
$ of solute molecules. Indeed, the whole Langevin formalism is
constructed in order to guarantee that the concentration fluctuations in
a small volume $\Delta V$ is
\begin{equation}
\label{fluctdiluted}
\left\langle \left( \Delta C_\nu \right) ^2\right\rangle
=\frac{C_\nu }{\Delta V},
\end{equation}
which is the equilibrium value for diluted solutions. In fact, for
a concentrated solution Eq. \ref{fluctdiluted} should be replaced
by
\begin{equation}
\label{fluct}
\left\langle \left( \Delta C_\nu \right) ^2\right\rangle =
\frac{1}{\Delta V}
\frac{T}{\left(\frac{\partial \mu}{\partial C} \right)_{P,T} }.
\end{equation}
Since the sidebranching amplitude in supersaturated solutions has been
found to be at least $1$ order of magnitude larger than predicted
in the diluted approximation, one concludes that it would be
necessary that the derivative of the chemical potential is $2$ orders
of magnitude greater than that given by the diluted approximation
$\frac{\partial \mu}{\partial C} \approx \frac{T}{C}$ in order to
explain the experiments by internal noise. Therefore, most likely
this internal noise is really not strong enough to account for the
observed sidebranching.
As we are not aware of quantitative data on thermodynamical
properties of supersaturated solutions that would permit us to improve
the estimations above, a definitive answer on the amplitude of
sidebranching in these dendrites remains open. In any case these
results call for experimental quantitative measurements in
solutal dendrites grown in diluted conditions, where Eq.
\ref{sidebrnoise} properly applies.

\section{Model and numerical procedure}

We have performed simulations of dendritic growth by employing a
phase-field model for solidification.
In this model both phases and their interface are
treated indistinctly,
and discriminated by
an effective non-conserved order parameter or phase field $\phi$,
which takes different values in each phase ($0$ and $1$ in our
simulations). This field changes smoothly across an interface
region of finite thickness, and its dynamics is coupled to that of
the diffusion field in such a way that the sharp-interface model
is recovered in the limit of vanishing interface thickness,
controlled by a new small parameter $\epsilon$. The equations of
the model read explicitly :
\begin{eqnarray}
\label{phasefield}
{\epsilon^2 \tau(\theta)} \frac{\partial \phi}{\partial t} =
\phi(1-\phi)
\left(\phi-{1 \over 2}+30 \epsilon \beta \Delta u \phi(1-\phi)\right)
\nonumber \\
-\epsilon^2 \frac{\partial}{\partial x} \left[ \eta(\theta)
\eta^{'}(\theta) \frac{\partial \phi}{\partial y} \right]
+\epsilon^2 \frac{\partial}{\partial y}
\left[ \eta(\theta)\eta^{'} (\theta)\frac{\partial \phi} {\partial
x} \right]
+\epsilon^{2} \nabla \left[ \eta^2(\theta) \nabla \phi \right]
\end{eqnarray}
\begin{equation}
\label{tempdiff}
{\frac{\partial u}{\partial t}}+{1 \over \Delta}
\left( 30 \phi^{2} - 60 \phi^{3} + 30 \phi^{4} \right)
\frac{\partial \phi}{\partial t} = \nabla^{2} u
+ \psi(x,y,t)
\end{equation}
where $u({\bf r},t) $ is the diffusion field and $\Delta$ is the
dimensionless undercooling. Lengths are scaled in some arbitrary
reference
length $\omega$, while times are scaled by $\omega^{2}/D$. In
these equations $\theta$ is the angle between the $x$-axis and the
gradient of the phase field. $\eta(\theta)$ is the anisotropy of the
surface tension. The anisotropy of the kinetic term is then given
by $\tau(\theta)/ \eta(\theta)$. $\beta$ is equal to
$\frac{\sqrt{2} \omega}{12 d_o}$ and $d_o$ is the capillary
length.

The external noise is introduced through the additive term $\psi$ in the
heat equation. This choice is not unique and is justified here only for
simplicity. Because of its external origin,
the noise is not assumed to satisfy a
fluctuation-dissipation relation. Furthermore, it is generally assumed
to be non-conserved,
as opposed to the more usual case of thermal noise, which
would enter the model equations as a stochastic current (i.e. conserved
noise) in the heat equation, and an additional stochastic term in the
phase-field equation \cite{karma2}.
In our simulations the noise term is evaluated at each uncorrelated cell
of lateral size $\Delta x$ simply as  $I \cdot r$,
where $I$ denotes the
amplitude of the noise, and $r$ is a uniform random number in the
interval $[-0.5,0.5]$.
The phase-field model equations have been solved
on rectangular lattices using first-order
finite differences on a uniform grid with mesh
spacing $\Delta x$.
An explicit time-differencing scheme has been used to solve the
equation for $\phi$, whereas for the $u$ equation the alternating-direction implicit (ADI) method was chosen,
which is unconditionally stable \cite{wheeler}.
The kinetic term has been taken as isotropic, which leads to
$\tau(\theta)=m\eta(\theta)$ with constant $m$. A four-fold  free
energy anisotropy $\sigma=\sigma(0)(1+\gamma cos(4\theta))$ has
been considered. This gives rise to dendrites growing with perpendicular side
branches. The values of $\gamma$ taken were
always smaller than $0.0625$, which ensured that we obtained rounded
shapes such as a parabola because forbidden directions were avoided
\cite{godreche}.

The growth morphologies were obtained by setting a small
vertical seed ($\phi=0$, $u=0$) in the centre of either of the two
shortest sides of the system and imposing
$\phi=1$ and $u=-1$ on the rest of the system. We have considered
symmetric boundary  conditions for $\phi$ and $u$ on the four sides
of the system, although they do not influence the results
presented in this paper.

We have used a set of phase-field model parameters that give rise
to a growing  needle without sidebranching for every anisotropy
$\gamma$ considered when no noise is added to the
simulations. This ensures that the sidebranching observed when
noise is present is not due to computational rounding. The fixed
phase-field model parameters  for all the simulations are
$\beta=400$, $m=20$ and $\epsilon=0.003$. The values of $\Delta$
and $\gamma$ have been varied in the range $0.5-0.7$ and
$0.045-0.060$, respectively. The noise amplitude is varied in the
range $5-150$. The time and spatial discretizations were kept
constant in all the simulations with values $\Delta x=0.01$ and
$\Delta t=10^{-4}$.  Two system sizes were used in the simulations.
A system with $1000 \times 1500$  grid points was used to observe fully
developed
sidebranching including tertiary arms and to have enough statistics
in order to compare  with measurements of the
sidebranching correlation at both sides of the dendrite. 
Additionally a
$500 \times 500$ grid points system was used when
the data sets did not require very extensive statistics. 
In Fig. 2 a growing dendrite is shown with
$\gamma=0.045$, $\Delta=0.6$, $I=11$ at a time $t=1.5$. The
velocity and the radius of the tip are very weakly affected when
noise is introduced. However, side branches appear at both sides
of the main dendrite, yielding approximately a $90^o$ angle, 
as was observed in \cite{dougherty}. Further down the tip
one can clearly observe competition between branches which gives
rise to a coarsening effect.  When branches reach
the vertical boundaries of the system, they are stopped by the
effect of the symmetric boundary conditions.
This did not affect the measurements presented in this paper, where
we have focused on the region between the tip and a point approximately
$150$ grid points down from the tip (grid points are marked on the
axes of Fig. 2). This region corresponds approximately to that
considered in the experimental work in Ref. \cite{dougherty} and,
moreover, it is an appropriate region in order to compare the behaviour of
sidebranching with the analytical results obtained from linear
perturbation theory. In the longer runs, and in order to avoid
working with unnecessarily large systems, we have performed
periodic shifts of the complete system. We have checked that this
did not affect the results of the simulation.

\section{Sidebranching characteristics}

In order to study the sidebranching induced by noise we have
measured the half-width $h_{z}(t)$ of the dendrite at various
distances $z$ behind the tip  as a function of time.
In order to have a comparable amount of data as in Ref.\cite{dougherty}
(they recorded around $240$ oscillations of the amplitude for each
$z$) we needed to simulate a dendrite four times longer ($t=6$)
than the one shown in Fig. 2.
The half width of the dendrite as a function of time at a distance
$z=40$ grid points from the tip  as well as its power spectrum
$P_{z}(f)$ are shown in Fig. 3. The same type of data corresponding
to a point
further from the tip ($z=100$ grid points) is shown in Fig. 4.  The
data used to compute the power spectra were
six times the lengths shown. Both sets of data have a strong
resemblance to those obtained experimentally \cite{dougherty}. We
also computed the cross-correlation function 
\begin{equation}
C(t^{'})=<[h_{Lz}(t+t^{'})-\bar{h}_{Lz}][h_{Rz}(t)-\bar{h}_{Rz}]>
/\sigma_{L} \sigma_{R},
\end{equation}
where $h_{Lz}(t)$, $h_{Rz}(t)$, $\sigma_{L}$ and $\sigma_{R}$ are
the half-width functions and their standard deviations for the two
sides of the dendrite at the same distance from the tip. We found
that $C(0)$ is approximately $0.4$ for points very close to the tip and
that its value decreases very quickly to $0$ when increasing $z$.
Moreover, in the points closer to the tip, $C(t^{'})$ drops to zero
after six oscillations, that is, the time to nucleate six side
branches. The same behaviour was observed in the experiments in
\cite{dougherty}. In simulations where we used a smaller data set,
the values of $C(t^{'})$ were larger than those found in the experiments,
as was to be expected.

Besides the predictions for the amplitude commented on above, the
behaviour of the wavelength is another main feature of 
sidebranching. It depends very weakly on $z$ \cite{brener2,karma2}:
$\lambda \sim z^{1/5}$. Despite the fact that 
experiments \cite{dougherty,qian} showed the predicted  dependence of the amplitude
on $z$, the same frequency of the
spectral peak for different $z$ was found.

In order to observe the  dependence of the sidebranching amplitude on $z$ in the simulations, the phase-field model
simulations were run on a $500 \times 500$ grid points system. In Fig. 5 we
show the square root of the area under the
spectral peak as a function of $z$. This representation gives the
amplitude behaviour of the sidebranching as a function of
the distance to the tip.
The data were obtained with $\gamma=0.045$, $\Delta=0.6$, $I=19$
until a time $t=0.5$.
The amplitude is found to
increase exponentially with $z^{2/5}$ for distances $z<100$.
Thus the behaviour of the data obtained in the simulations is
consistent with the linear analysis \cite{brener2,karma2} up to a
certain value of $z$.
In \cite{dougherty}
similar behaviour was found in the linear regime, although they
found a saturation of the amplitude further down from the tip. This
could be due to some boundary effects which stop the growth of the
side branches.

The dependence of the sidebranching wavelength on the distance to
the tip was studied with the data recorded in the same simulation
as for Fig. 5 and after performing the power spectrum. The wavelength 
was found to remain constant in the
interval of considered $z$, which is consistent with the observations in
\cite{karma2}.

\section{Concluding remarks}

We have studied the sidebranching induced by fluctuations in
dendritic growth, for which there is good quantitative agreement
between theory and experiment for thermal dendrites (i.e.
controlled by heat diffusion). This agreement has only been
qualitative for solutal dendrites, as there was no prediction
available for their sidebranching amplitudes.
In order to be able to perform a quantitative comparison,
we have obtained an estimation of sidebranching
amplitudes originated by internal fluctuations in solutal diluted
dendrites. The resulting values appear as much smaller (at least one
order of magnitude) than those observed in some experiments (performed in
concentrated solutions). This can be attributed to the effect of
other noise sources, of an external origin. To confirm this
conclusion, it would be necessary to make use of quantitative
experimental results obtained in more diluted conditions.

We have obtained the
effects of non-internal fluctuations on dendritic sidebranching by
introducing  non-conserved noise in a phase-field model for
solidification.
Our simulations qualitatively reproduce well previous
experimental and analytical results. In particular, we have
reproduced
the dependence of the sidebranching amplitude 
on the distance from the tip, confirming the validity
of what was previously obtained with a linear theory including
internal noise.
Thus, the qualitative behaviour of the 
sidebranching amplitude, when this
is due to the selective amplification of fluctuations, is basically
independent of the origin of the noise. In conclusion, qualitative
concordance between experimental results does not directly imply a
common source of fluctuations and therefore, a careful quantitative
study of sidebranching activity may help to elucidate the origin
of the dominant noise in each experiment.

The phase-field model appears to be a versatile method to study
dendritic growth in general and sidebranching characteristics in
particular. It has been shown to be adapted to take into
account thermodynamical fluctuations \cite{pavlik,karma2,pavlik2}, but
alternative ways to introduce noise (such as the one employed in this work)
appear to be appropriate to qualitatively reproduce the behaviour
of sidebranching when the noise is of an external origin.

Our results on sidebranching can also
be helpful for further simulations where qualitatively realistic
sidebranching needs to be distinguished from that generated by
numerical noise due to the round-off of the computer.

\section{Acknowledgements}

The authors wish to thank Prof. W. Van Saarloos and Dr. T. Toth Katona
for many fruitful discussions. R.G.C. is indebted to the Lorentz
Institut for its kind hospitality and assistance.
The authors thank the Direcci\'on General de
Investigaci\'on Cient\'{\i}fica y T\'ecnica (Spain) (Projects
BFM2000-0624-C03-02, BFM2000-0628-C03-01 and BXX2000-0638-C02-02),
Comissionat per a Universitats i Recerca (Spain) (Projects
1999SGR00145 and 1999SGR00041), and European Commission
(TMR Network Project ERBFMRXCT96-0085)
for financial
support. We also acknowledge the Centre de Supercomputaci\'o de
Catalunya (CESCA) for computing support.

\newpage
\section{Figure Captions}

\noindent
FIG. 1. Amplitude vs. distance to the tip $z$ of the 
sidebranching induced by internal
noise in the experiments of Ref. \cite{dougherty}.
Line: theoretical prediction of Eq. \ref{sidebramp}
with a overestimated value 
of $\overline{S}$;
crosses:
experimental results (taken with permission 
from Ref. \cite{dougherty}).
Quantities are expressed in micrometers.
\\
\noindent
FIG. 2. Dendrite obtained with the phase-field model with a noise
term included, as is described in the text. Ticks denote number of grid points.
\\
\noindent
FIG. 3. Half-width $h_{z}(t)$ of the dendrite and its power
spectrum $P_{z}(f)$ ($P$ in square grid points times dimensionless time and 
frequency in cycles per dimensionless time) for $z=40$ grid points.
\\
\noindent
FIG. 4. Half-width $h_{z}(t)$ of the dendrite and its power
spectrum $P_{z}(f)$ ($P$ in square grid points times dimensionless time and
frequency in cycles per dimensionless time) for $z=100$ grid points.
\\
\noindent
FIG. 5. Square root of the area under the spectral peak as a
function of $z$ (both in grid points). The line indicates exponential growth
with $z^{2/5}$.\\
\noindent
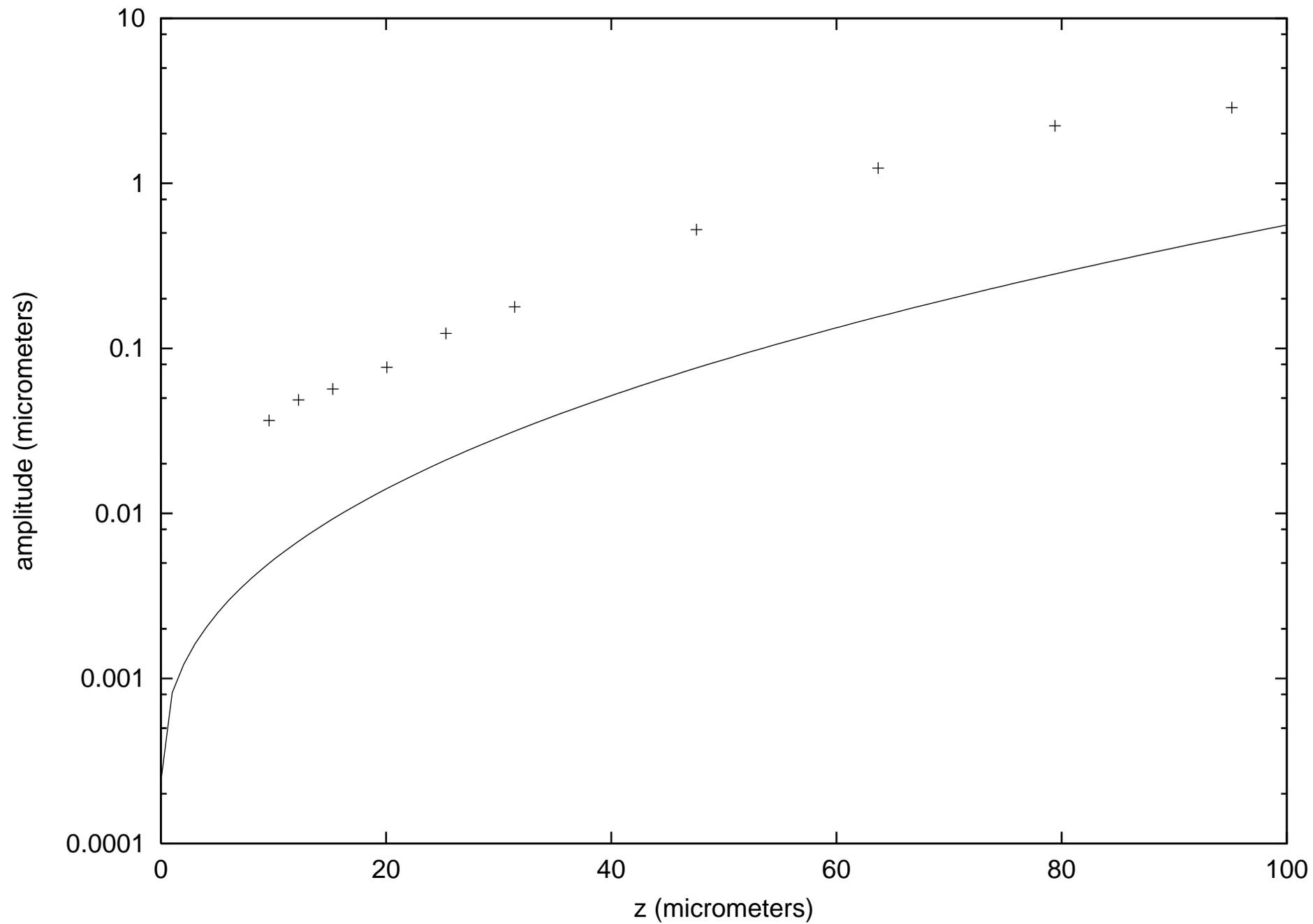
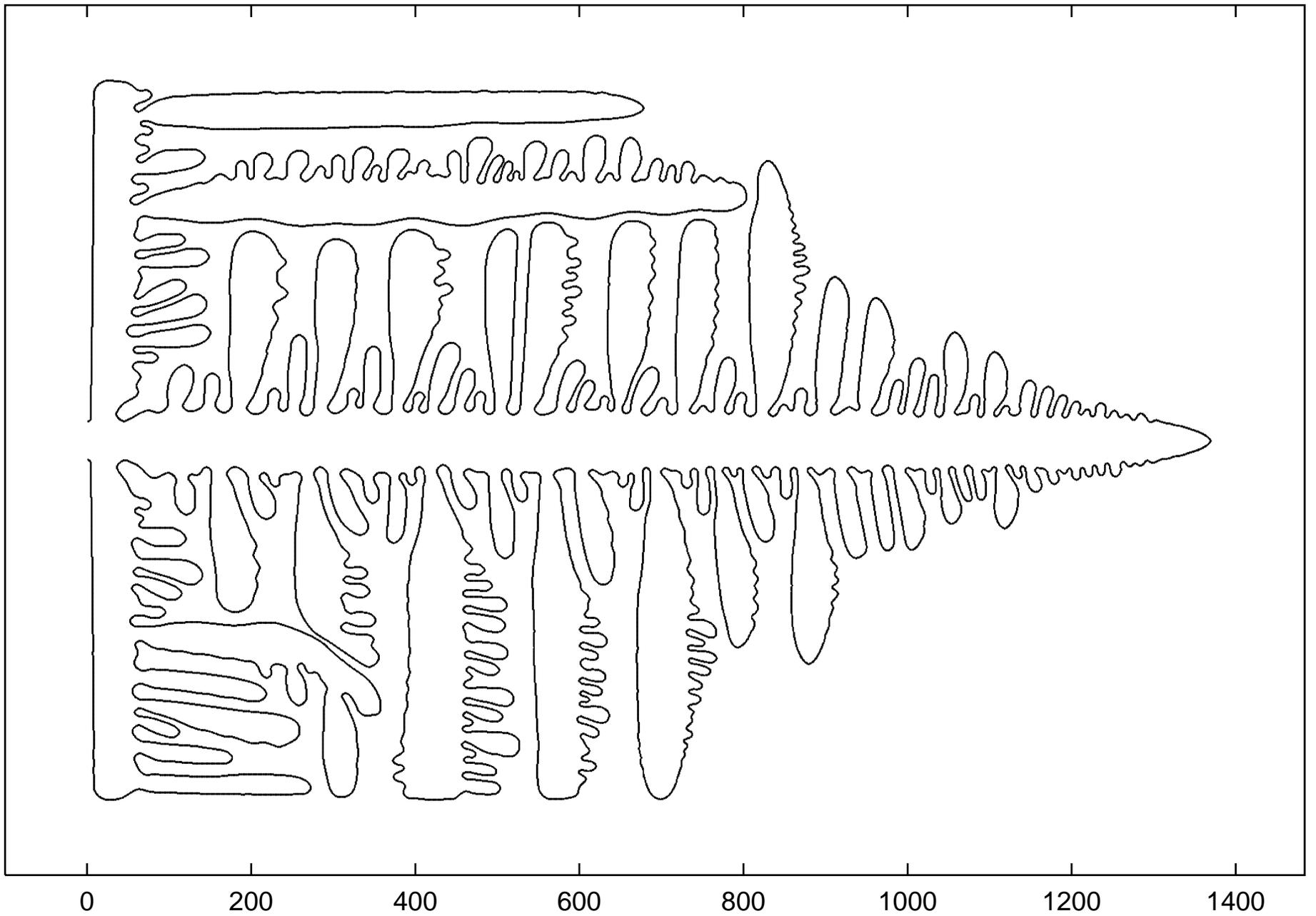
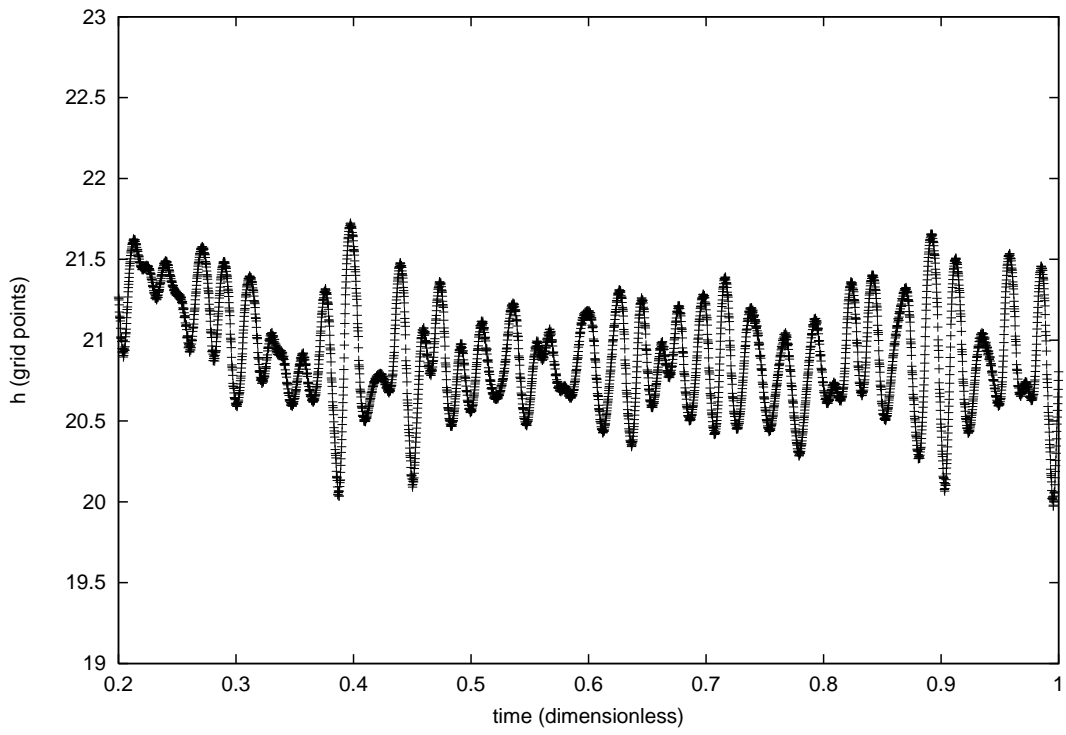
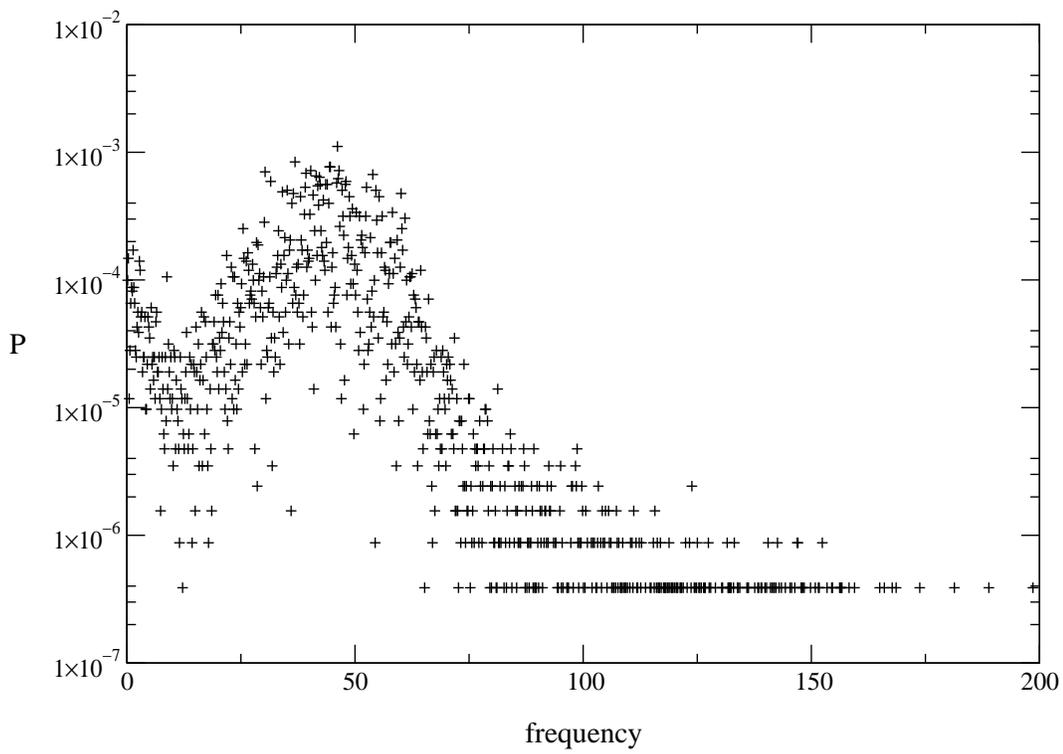
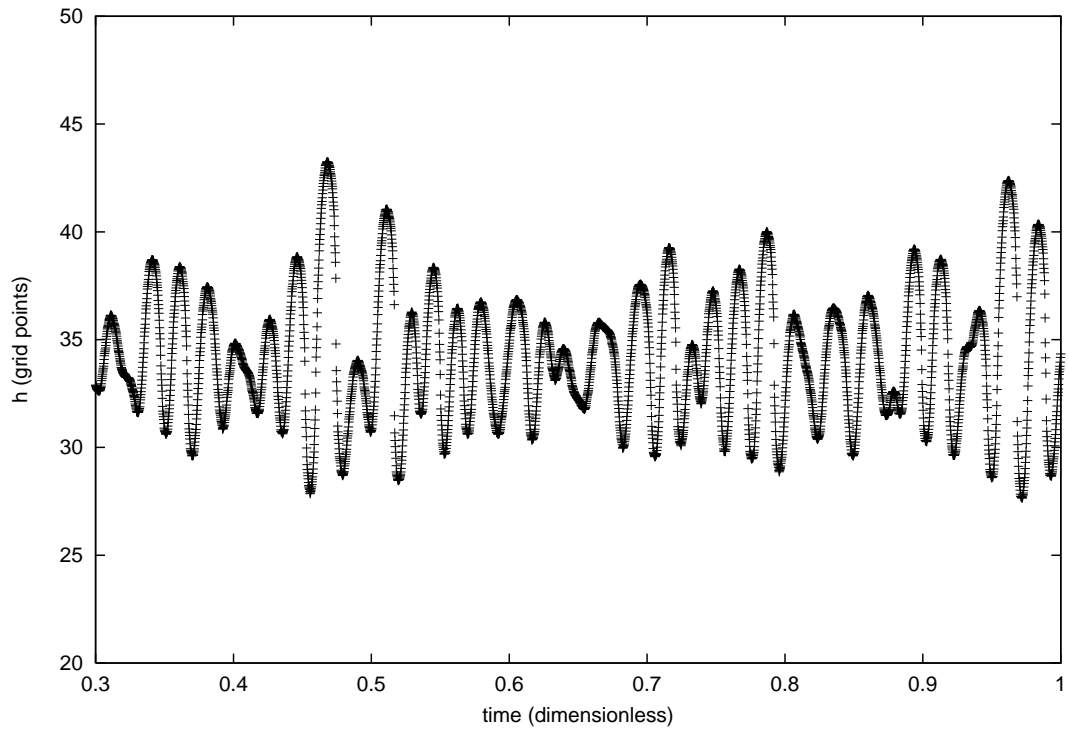
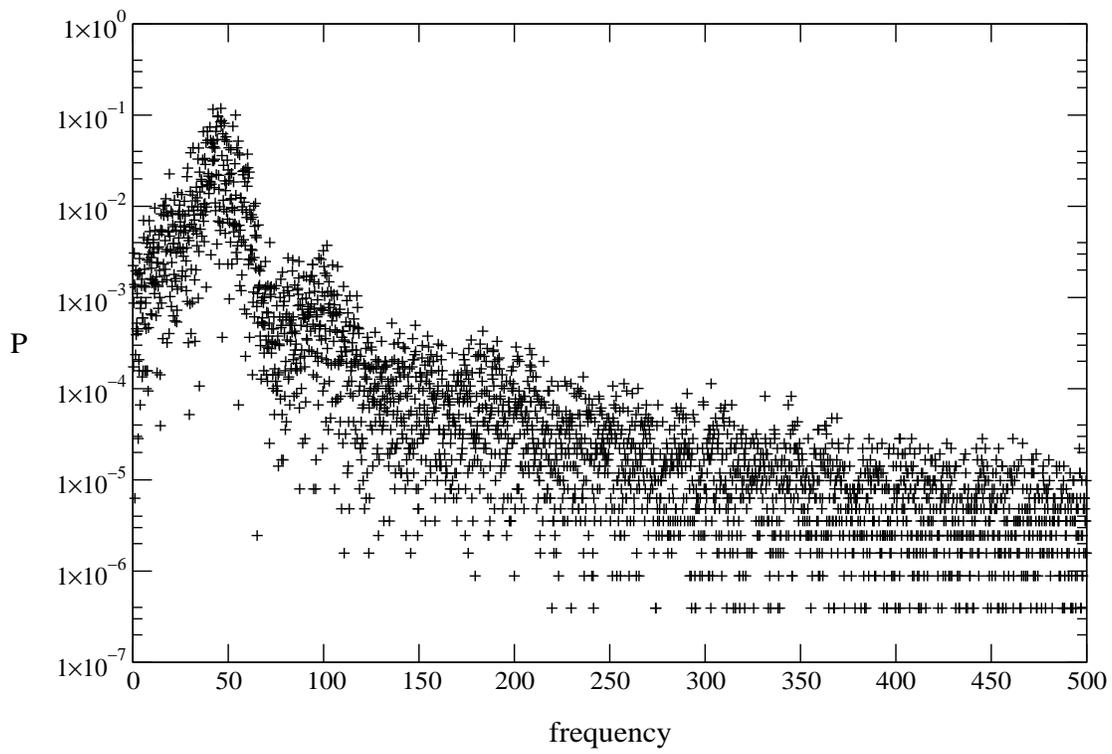
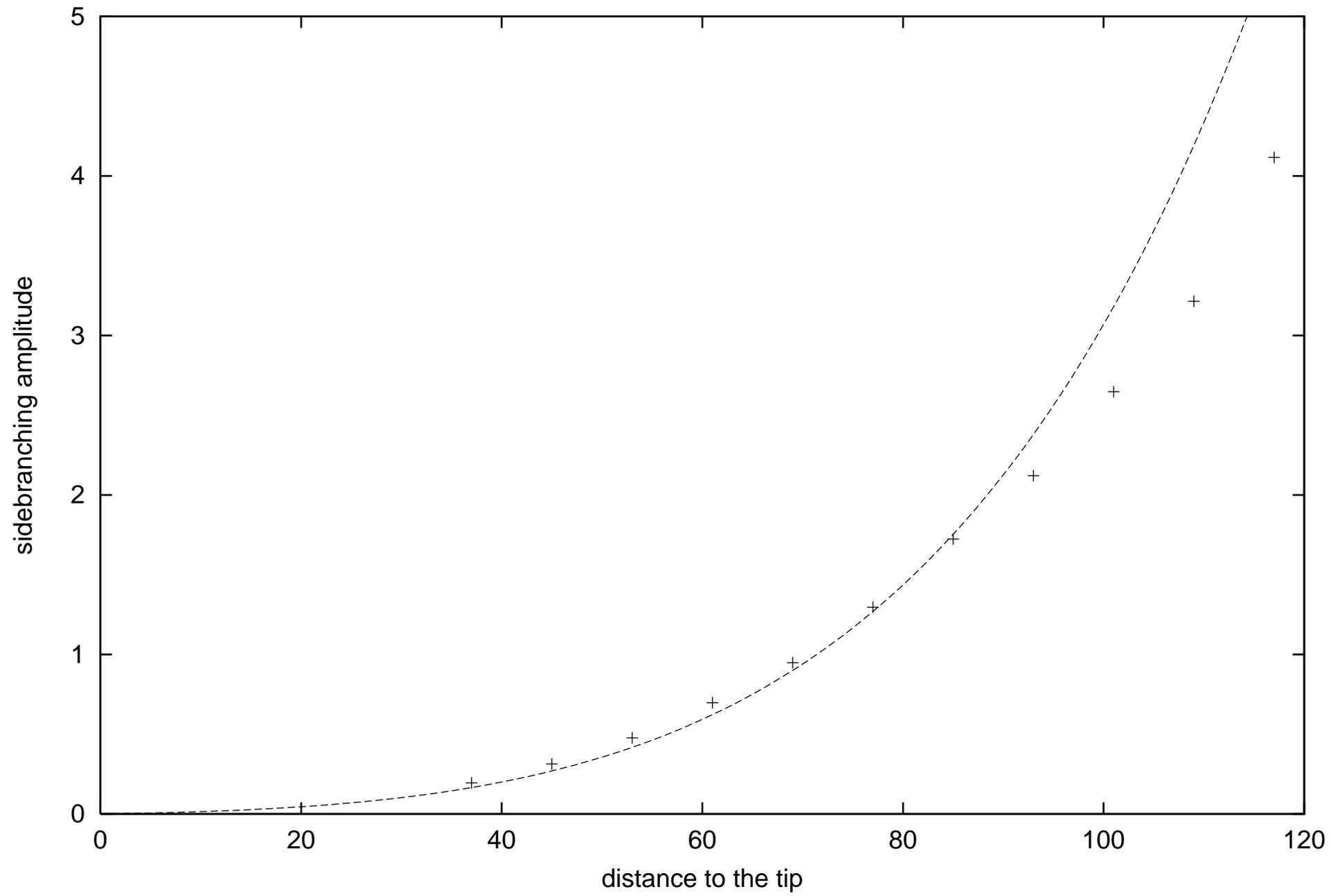
\end{document}